# Hardware-Aware Characterization of Edge AI Inference under LLM-Driven Fault Injection

1st Faezeh Pasandideh
*Department of Electrical Engineering*
*Hamm-Lippstadt University of Applied Sciences (HSHL)*
Lippstadt, Germany
faezeh.pasandideh@hshl.de

2nd Mehdi Azarafza
*Department of Electrical Engineering*
*Hamm-Lippstadt University of Applied Sciences (HSHL)*
Lippstadt, Germany
mehdi.azarafza@hshl.de

3rd Achim Rettberg
*Department of Electrical Engineering*
*Hamm-Lippstadt University of Applied Sciences (HSHL)*
Lippstadt, Germany
achim.rettberg@hshl.de

***Abstract*—As deep learning models are deployed on resource-constrained edge platforms in autonomous driving systems, reliable knowledge of hardware behaviour under input degradation becomes an essential requirement. We presented a systematic characterisation of GPU utilisation, RAM consumption, power draw, throughput, tail latency, and thermal behaviour of TensorRT-optimised YOLOv10s, YOLOv11s, and YOLO2026n pipelines on an NVIDIA Jetson Nano under a large-scale fault-injection campaign across lane-following and object detection tasks, using faults synthesised via a decoupled LLM–LDM framework from our JetBot platform data. All three pipelines hold up well under heavily degraded inputs: GPU occupancy remains stable at 23–24%, memory settles into a consistent release pattern after warm-up, thermal rise stays within safe bounds, and real-time throughput is maintained across all conditions. YOLOv10s is the most power-efficient but lags in retention and tail latency; YOLOv11s leads in throughput; and YOLO2026n delivers the best overall balance of retention (92.4%), tail-latency stability, and task-invariant behaviour, making it the strongest candidate for safety-critical edge deployment. These results offer a hardware-level reliability baseline that motivates future fault-aware training strategies for autonomous edge AI systems.**

## I. Introduction

Deploying AI vision systems on resource-constrained edge platforms for autonomous driving applications introduces a dual challenge: maintaining inference accuracy under real-world degraded conditions while keeping hardware resource consumption within safe operational bounds. Although significant progress has been made in optimizing deep learning models for edge deployment, the behavior of these models at the hardware level when inputs are corrupted by sensor failures, lighting degradation, or adversarial visual conditions remains poorly understood.

Existing studies largely evaluate inference pipelines under normal operating conditions, treating resource monitoring as a performance optimization tool rather than a reliability indicator. Fault detection work tends to focus on timeout-based or software-level failure signals, without profiling the accompanying CPU load, GPU utilization, RAM consumption, or power draw during degraded inference. This leaves a gap in our understanding of how edge-deployed models actually behave when things go wrong.

To address this, we employ a two-phase decoupled framework [1] in which large language models (LLMs) and latent diffusion models (LDMs) generate synthetic fault scenarios offline, producing thousands of degraded input folders from data originally collected by our JetBot platform. These faults simulate realistic sensor and environmental degradations representative of autonomous driving conditions. The proposed framework is shown in Figure 1. In contrast to prior work, our approach focuses on hardware-level characterization of inference behavior under fault injection, rather than solely on model accuracy or detection performance. In the online phase, the generated dataset is used to drive the TensorRT-optimized YOLOv10, YOLOv11s, and Yolo2026v2 pipelines on NVIDIA Jetson Nano, while CPU load, GPU utilization, RAM consumption, power draw, throughput, and thermal behavior are systematically recorded across both lane-following and object detection tasks.

The main contributions of this work are as follows:

- A hardware-aware evaluation of edge AI inference under semantically generated fault scenarios;
- A comprehensive analysis of resource utilization, throughput, and thermal behavior under degraded visual inputs;
- An empirical study across both lane-following and object detection pipelines on a resource-constrained edge platform.

The subsequent sections of this paper are organized as follows. Section II reviews related work on fault injection for neural networks, edge AI deployment, and hardware resource characterization. Section III describes the system design and implementation, covering the JetBot platform, the TensorRT-optimized YOLOv10, YOLOv11s, and YOLO2026v2 pipelines, and the NVIDIA Jetson Nano deploy-ment setup. Section IV details the dataset collection procedure and the preparation of the VisionFault-920K fault-injected image corpus. Section V presents the decoupled fault injection framework, including the LLM-based fault scenario generation and the LDM-based image synthesis pipeline. Section VI reports the evaluation results, analyzing ΔPower, ΔRAM, ΔTemp, GPU%, ΔFPS, latency, and detection retention across object detection and lane-following tasks under fault injection. Finally, Section VII concludes the paper and outlines directions for future work.

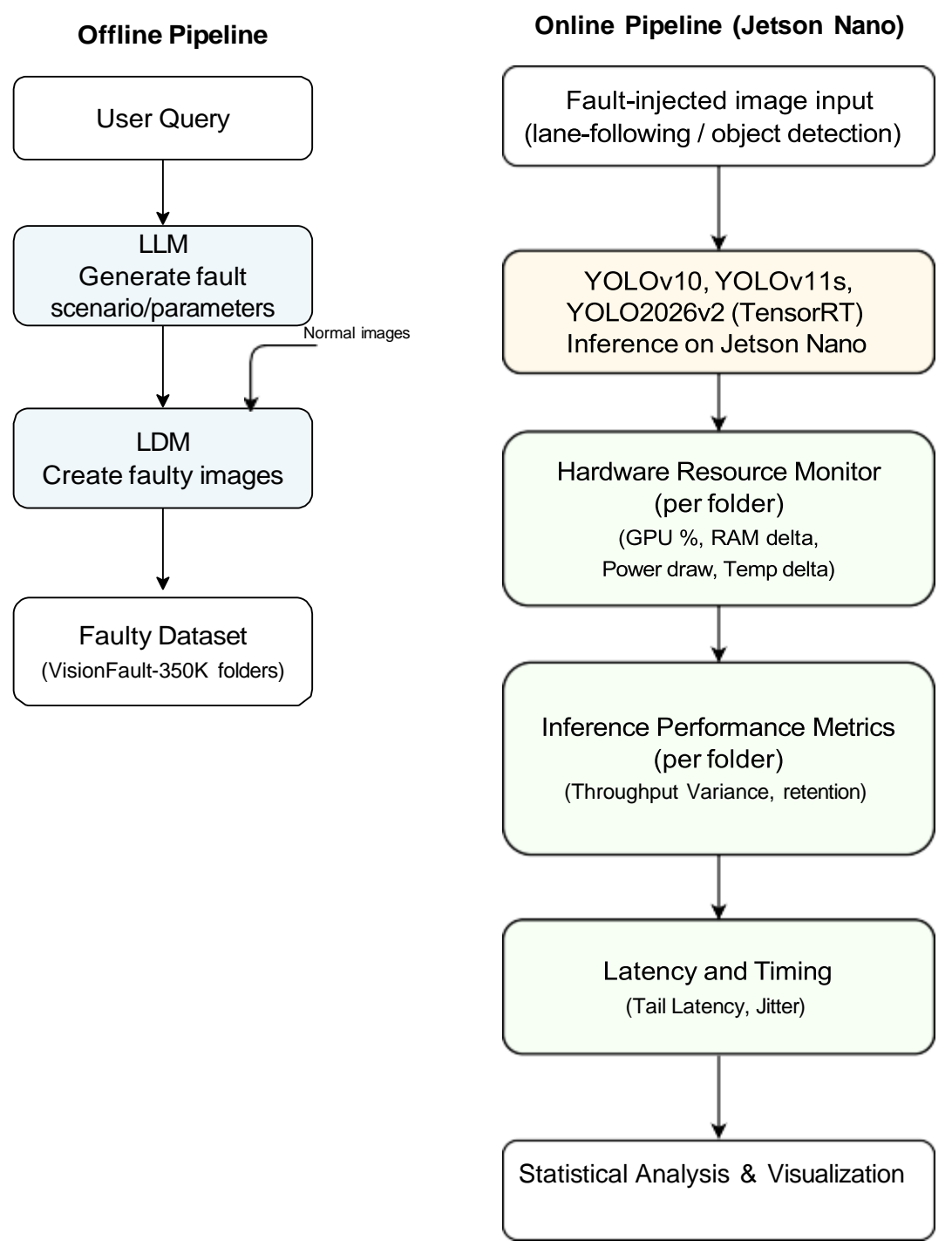

Fig. 1: Two-phase pipeline: offline synthetic fault generation via LLM-LDM; online hardware characterization under fault injection on Jetson Nano.

## II. Related Work

Authors in [2] conducted a systematic empirical study of timeout-based fault detection under controlled resource stress on heterogeneous edge platforms, evaluating six representative workloads, including CPU/GPU inference using YOLOv3, across Raspberry Pi 4B and Jetson Nano devices under five stressors: CPU overload, memory contention, disk I/O, cache thrashing, and page faults. Their results demonstrated that static timeout calibration is fundamentally brittle at the edge, with tight thresholds producing up to 30% false positives in fault-free runs, while memory contention consistently emerged as the dominant stressor driving timeout misclassifications across both platforms. However, their work focuses exclusively on timeout-based failure detection rather than broader hardware-level behavior during inference under fault injection, and does not monitor fine-grained resource metrics such as GPU load, power consumption, or memory delta during model inference.

Authors in [3] proposed a concurrent multi-frame processing scheme for real-time object detection on resource-constrained edge devices, grouping video frames according to the number of available CPU cores and distributing them across cores for parallel, synchronization-free inference using YOLO models deployed on an NVIDIA Jetson Orin Nano. Their evaluation across multiple real-world video and image datasets, including MS-COCO, ImageNet, and PascalVOC, demonstrated runtime improvements of up to 4.45×, memory reduction of up to 69%, and power consumption decrease of up to 73% per frame compared to serial execution. However, their work focuses exclusively on throughput and resource efficiency optimization through parallelism, and does not investigate the impact of fault injection on inference behavior or hardware utilization. Furthermore, their resource monitoring serves as a performance optimization metric rather than a reliability-oriented evaluation, and no analysis is provided on how the system behaves under degraded or adversarial input conditions.

Authors in [4] conducted a systematic performance benchmarking of multiple YOLOv8 variants on the NVIDIA Jetson Orin NX platform, evaluating inference latency, frame throughput, and computational resource usage, including GPU memory and RAM, across varying input resolutions and batch sizes under both PyTorch and TensorRT-optimized deployments. Their findings demonstrated that the TensorRT engine format outperforms PyTorch by 17.7% at batch size 2, while GPU memory scales linearly with batch size (69% increase from batch 1 to 4), and consistent Out-of-Memory (OOM) failures were observed when TensorRT exceeded batch sizes of 8, identifying critical operational stability limits on resource-constrained edge hardware. However, their work focuses exclusively on performance benchmarking under normal inference conditions, without introducing any fault injection or input degradation scenarios to evaluate model robustness. Furthermore, while GPU memory and RAM usage are monitored, CPU utilization and power consumption are explicitly acknowledged as future work, leaving a gap in comprehensive hardware-level characterization during degraded inference.

Authors in [5] proposed an end-to-end edge-deployed adaptive traffic light control system that integrates YOLO-based vehicle detection with a *Passenger Car Equivalent* (PCE)-aware traffic density estimation module, deployed on a Jetson Xavier NX road-side edge node (RSEN) achieving 90% mAP and 74 FPS under real-world undisciplined traffic conditions in Karachi, Pakistan. Their system demonstrated that lightweight models such as YOLOv8-nano (6 MB) and YOLOv7-tiny (11.8 MB), optimized via TensorRT, can sustain real-time inference on resource-constrained Jetson Nano and Xavier NX platforms, reducing intersection-level congestion by up to 33% and average waiting time by 23% compared to fixed-time baselines. However, their work evaluates model performance exclusively under normal operating conditions, with no investigation of how synthesized or injected visual faults, such as sensor degradation, lighting failures, or adversarial input corruptions, affect inference accuracy, hardware-level resource utilization (CPU load, RAM consumption, power draw), or control-level decision quality on edge hardware. Furthermore, while GPU memory is implicitly constrained by the Jetson platform, no systematic power consumption profiling is conducted during inference, an aspect explicitly deferred to future work by the authors.

Taken together, these gaps highlight the need for a more reliability-oriented and hardware-aware perspective that complements existing performance-focused studies. Our work contributes in this direction by characterizing CPU load, GPU usage, RAM consumption, and power draw of the TensorRT-optimized YOLOv10, YOLOv11s, and YOLO2026v2 pipelines on NVIDIA Jetson Nano under systematic fault injection across both lane-following and object detection scenarios, where original data collected by our JetBot platform is subsequently corrupted by synthetically generated faults produced through a combination of LLM and LDM, offering a hardware-level understanding of model behavior under degraded input conditions.

## III. System Design and Implementation

### A. Robotic Platform and Operating Environment

The experimental platform is a custom-built autonomous mobile robot based on the NVIDIA Jetson Nano [6] edge-computing device. The robot has a compact chassis with dimensions of approximately 15 cm × 15 cm × 20 cm, measured axle-to-axle and vertically to the camera mount. The system is equipped with an RGB camera operating at 1280×720 resolution at 30 FPS with fixed focus and 55° diagonal field of view [7], mounted at the front of the robot with a 45° downward viewing angle to ensure continuous visibility of the navigation path and obstacles. The robot performs line following using visual input, while simultaneously executing onboard deep-learning-based object detection for obstacle awareness. The navigation line is tracked visually, and obstacles are detected in the forward path. The robot operates at a variable linear velocity between 0.3 m/s and 0.8 m/s depending on curvature and obstacle proximity, and is designed for continuous operation of up to 8 hours, requiring stable long-term inference and control behavior. The navigation controller supports straight-line tracking, sharp turns, and obstacle-triggered braking and avoidance.

Experiments were conducted in an indoor warehouse environment with concrete gray flooring, occasional reflective surfaces, and a black tape navigation line of 5–10 cm width. Lighting conditions combined natural daylight from west-facing windows and overhead fluorescent lighting, with recordings spanning 06:00 to 18:00 and noticeable sun glare between 14:00 and 16:00. Additional environmental factors included dust accumulation, illumination variation, temperature changes (15°C to 30°C), and intermittent indoor–outdoor transitions, introducing realistic operating variability.

### B. Decoupled Perception and Navigation Architecture

The robot's perception and navigation subsystems are fully decoupled and executed concurrently. Both modules receive frames from a shared camera acquisition process, but operate independently without exchanging control signals. The line-following module runs in a daemon thread controlling motors via GPIO and ultrasonic sensor input, while the deep learning inference module runs in the main thread for CUDA compatibility, processing frames for object detection and visualization.

This design isolates perception faults from control logic, enabling controlled analysis of inference behavior without propagating errors to actuation. Such decoupling is essential for fault-injection studies, as it allows repeatable and fine-grained evaluation of perception-layer robustness independent of navigation dynamics.

### C. Vision Pipeline and Object Detection

Visual data from the front-facing camera is processed by a deep learning–based object detection module implemented as a standalone runtime in the main thread to satisfy CUDA context requirements. The module loads YOLOv10, YOLOv11s, and YOLO2026v2 TensorRT-optimized engines and performs inference on incoming frames. Input frames are resized and normalized to the model's expected resolution, and bounding boxes, confidence scores, and class predictions are extracted and post-processed using non-maximum suppression to remove duplicate detections. The module configuration, including batch size, confidence thresholds, minimum box area, top-k selection, and Non-Maximum Suppression (NMS) parameters, is selected to balance throughput and detection stability on the embedded platform. Detected objects are rendered on a copy of the frame with class labels and confidence scores, and video output is stored for offline analysis. The inference module does not influence robot motion, ensuring separation between perception evaluation and control execution.

### D. Line-Following and Control Logic

The robot follows a black navigation line using a vision-based line detection module executed in a separate worker thread. Frames are shared via a synchronized buffer, allowing the line-following module to access the latest frame independently of the inference process. Frames are converted to HSV color space, and a black color mask is applied, followed by morphological operations for noise reduction. A region of interest covering the bottom 40% of the frame is extracted, and the centroid of the masked region is computed. The lateral deviation from the image center determines wheel control: large deviations trigger hard left or right turns, while smaller deviations are corrected using proportional control through motor PWM adjustments. Obstacle avoidance is handled independently using an HC-SR04 ultrasonic sensor, and motors are stopped if an obstacle is detected within a 4 cm threshold. Similarly, if the line is lost, the robot halts until reacquisition. This control design ensures stable navigation while allowing independent evaluation of perception under degraded inputs.

## IV. Dataset Collection and Preparation

### A. Data Collection

All image data used for training and evaluation were collected directly using the autonomous mobile robot platform described in Section III-A. A forward-facing RGB camera operating at 320×240p and 20 FPS was mounted at a downward angle of approximately $45^{\circ}$. Video recordings were captured while the robot followed a black floor line in an indoor warehouse environment under varying lighting conditions. A custom Python script was developed to extract every single frame from the recorded video sequences without temporal down-sampling. From this full frame set, frames were manually selected to remove redundant and visually uninformative samples while preserving diversity in illumination, background clutter, and partial occlusions, resulting in 1489 RGB images (224×224), available on GitHub. The mobile robot platform, equipped with a front-mounted camera module on a servo gimbal, four-wheel drive chassis, and onboard electronics for edge inference and fault injection experiments is shown in Figure 2.

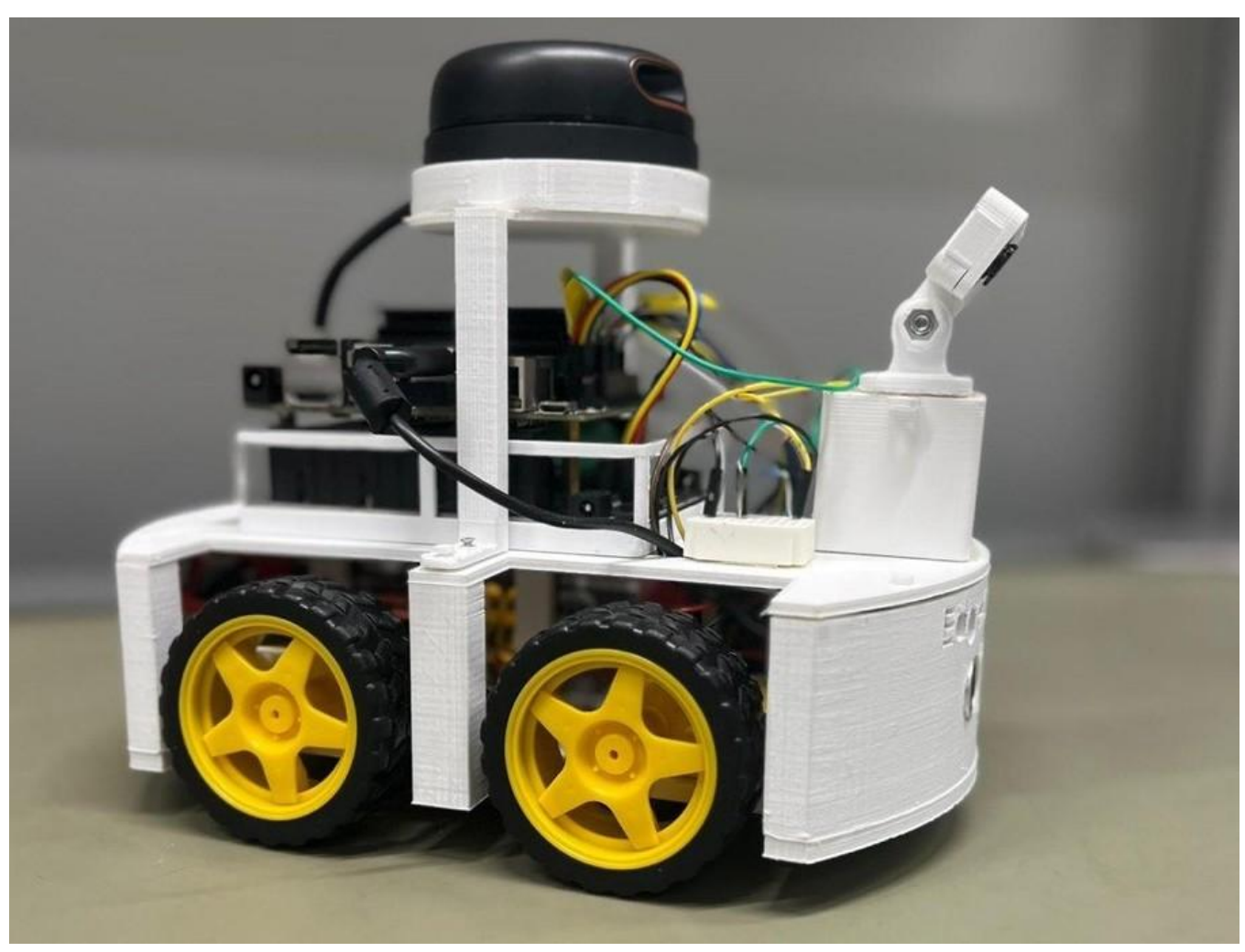

Fig. 2: The robot platform used for edge inference experiments.

Due to the nature of the mobile robot's navigation strategy, the robot followed a single black guide line in one direction only throughout all recording sessions. As a result, all objects were observed from a consistent dominant viewpoint, and only lighting conditions were intentionally varied. This design choice ensured that viewpoint variability was controlled, allowing the experimental analysis to primarily isolate the effects of fault injection and environmental illumination on object detection behavior rather than geometric pose variation. The primary object categories included animals, cars, people, and traffic signs.

### B. Model Training and Embedded Deployment

Model training was performed using a cloud-based Google Colab environment with a Python 3 runtime and an NVIDIA T4 GPU accelerator [8]. The object detection network was trained using YOLOv10, YOLOv11s, and YOLO2026v2 architectures [9] with the PyTorch [10] framework, with an input image resolution of 256×256, a training duration of 180 epochs, and single-GPU execution. The selection of YOLO was motivated by its lightweight architecture and suitability for real-time inference on embedded GPU platforms [11], particularly the NVIDIA Jetson series. After training, model weights were exported in PyTorch (`.pt`) format, subsequently converted to ONNX [12] format using the Ultralytics library, and finally compiled into a TensorRT [13] engine directly on the NVIDIA Jetson platform for real-time inference during fault-injection testing.

## V. Fault Injection Framework

As illustrated in Fig. 1, the proposed architecture enables a clear separation between offline fault generation and online inference execution, allowing controlled hardware-level characterization under systematically injected degradation scenarios. The central objective of this work is to evaluate object detection robustness under fault injection, with a focus on lane following. The injected faults are software-based, designed to avoid permanent hardware damage, and adhere to principles of randomized activation timing, reversibility after each test run, and controlled fault duration, ensuring repeatable and safe evaluation of perception-layer performance.

To ensure comprehensive coverage, fault scenarios were generated iteratively using GPT-OSS 120B [14] with Ollama [15], prompting the model to produce 100 scenarios per iteration, ultimately yielding 7,950 valid scenarios. A subsequent cleaning stage was applied to remove any JSON entries found to be malformed, incomplete, or improperly structured, ensuring only well-formed scenarios were carried forward. A Latent Diffusion Model (Stable Diffusion 2.1 [16]) then serves as the core degradation mechanism, using the LLM's textual output to synthesize degraded images via an Img2Img process, where the Denoising Strength parameter controls the degree of image corruption. Our final faulty dataset, VisionFault-350K [17], contains 350,751 fault-augmented images derived from real robotic camera recordings, covering primary tasks of lane following and obstacle detection, spanning diverse fault categories including camera failures, motion blur, extreme weather (ice, rain, fog), low light conditions (tunnel, night, backlight), and lens distortions. The dataset is publicly available on Zenodo. Figure 3 shows randomly selected examples of normal images alongside faulty samples generated by the LDM and LLM-based fault injection framework.

## VI. Evaluation and Results

Our evaluation systematically investigates hardware resource utilization under fault-injected conditions by assessing three TensorRT-optimized engines deployed on the NVIDIA Jetson Nano: YOLOv10s, YOLOv11s, and YOLO2026n. For each model, three inference runs were performed: one for the lane-following task (1,024 folders, 129 images per folder) and one for the object detection task (1,642–1,645 folders, 129 images per folder, totaling up to 212,205 images).

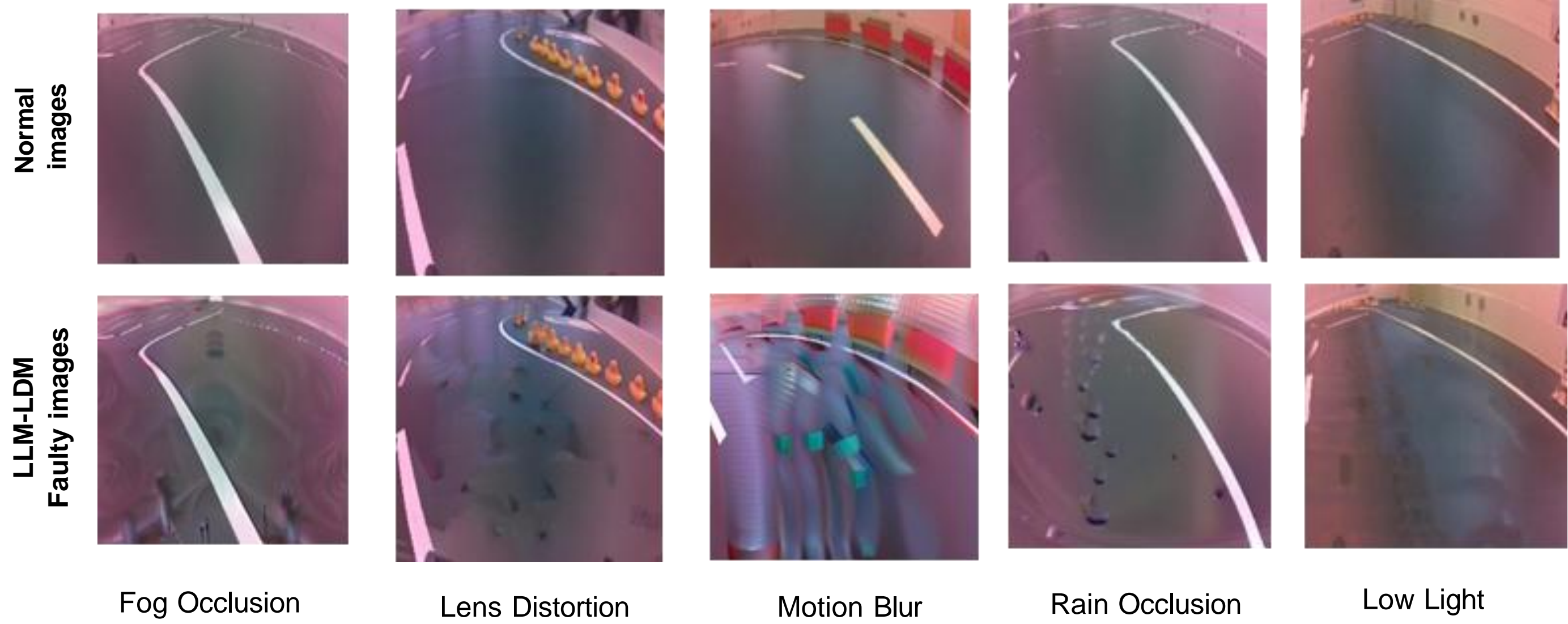

Fig. 3: Normal and synthetically generated faulty image samples

### A. Hardware Resource Utilization

We analyze the distributions of key hardware metrics, including Power Draw Variance ($\Delta$Power), Memory Allocation ($\Delta$RAM), Thermal Flux ($\Delta$Temp), and Core Utilization (GPU%), for both object detection and lane-following tasks. All measurements are aggregated across the complete set of fault-injected folders on the NVIDIA Jetson Nano, enabling a comprehensive assessment of performance under systematically degraded input conditions.

*1) Power Draw Variance ($\Delta$Power):* The `Power_Delta` distributions show the change in power consumption (W) under fault injection for lane following and object detection, compared to baseline runs for YOLOv10s, YOLOv11s, and YOLO2026n (Figs. 4 and 5).

YOLOv10s exhibits near-zero, tightly centred distributions: mean $-0.010$ W ($\sigma$ = $0.107$ W) for lane following and $+0.011$ W ($\sigma = 0.095$ W) for object detection, with operational bands $[-1.16, +0.31]$ W and $[-0.975, +0.417]$ W, confirming negligible and consistent power overhead.

YOLOv11s shows Gaussian-like distributions peaking around $0.8$–$1.0$ W. Lane following yields mean $+0.874$ W ($\sigma = 0.131$ W) and object detection $+0.894$ W ($\sigma = 0.116$ W), with occasional low-power outliers and operational bands extending to $\sim 1.3$–$1.4$ W. Absolute mean total power is $5.199$ W and $5.220$ W, respectively.

YOLO2026n has the highest, right-shifted distributions: lane following mean $+1.414$ W ($\sigma = 0.114$ W), object detection $+1.415$ W ($\sigma = 0.107$ W), with operational bands $[+0.140, +1.830]$ W and $[+0.430, +1.830]$ W, confirming consistent power increase across all runs.

Overall, the power-draw hierarchy is YOLOv10s $\approx 0$ W $<$ YOLOv11s $\approx +0.88$ W $<$ YOLO2026n $\approx +1.41$ W, reflecting increasing model complexity. All models maintain stable power under extended fault injection.

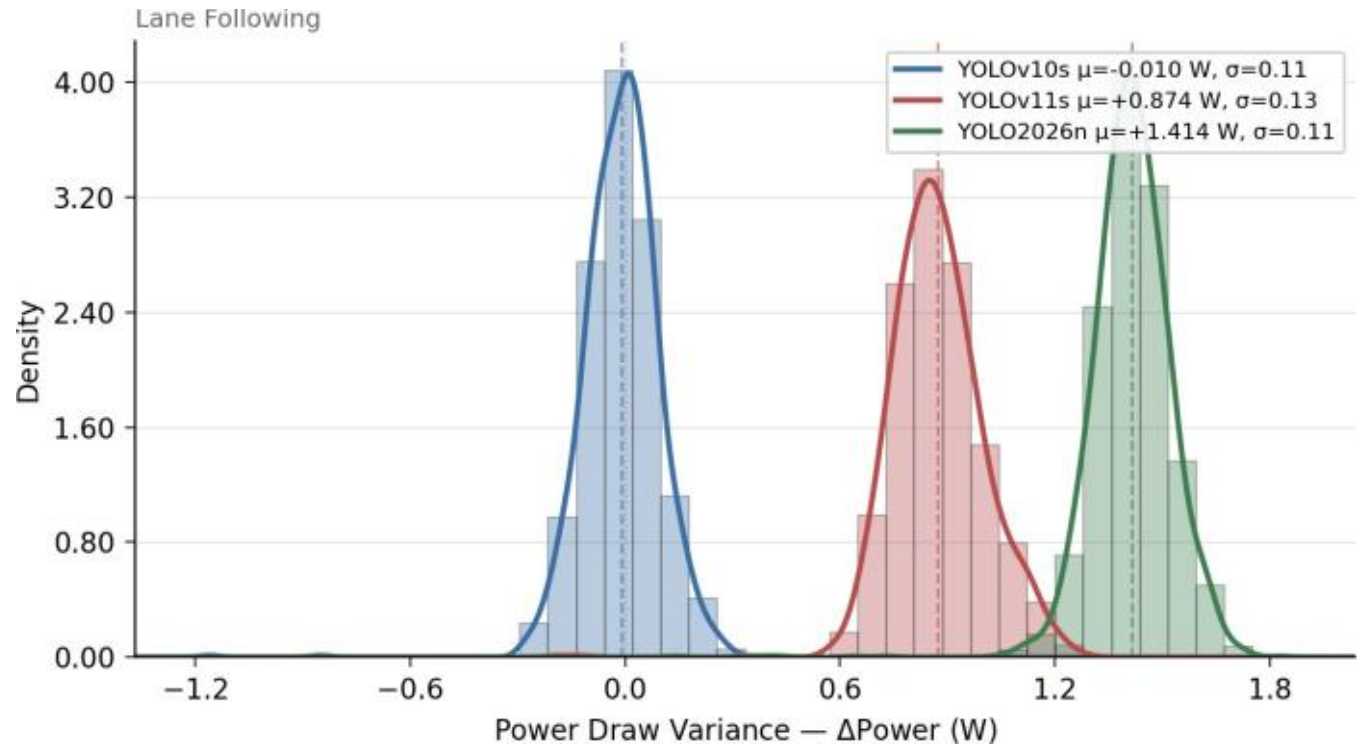

Fig. 4: Power Draw Variance ($\Delta$Power) for YOLO models under fault injection — lane-following task.

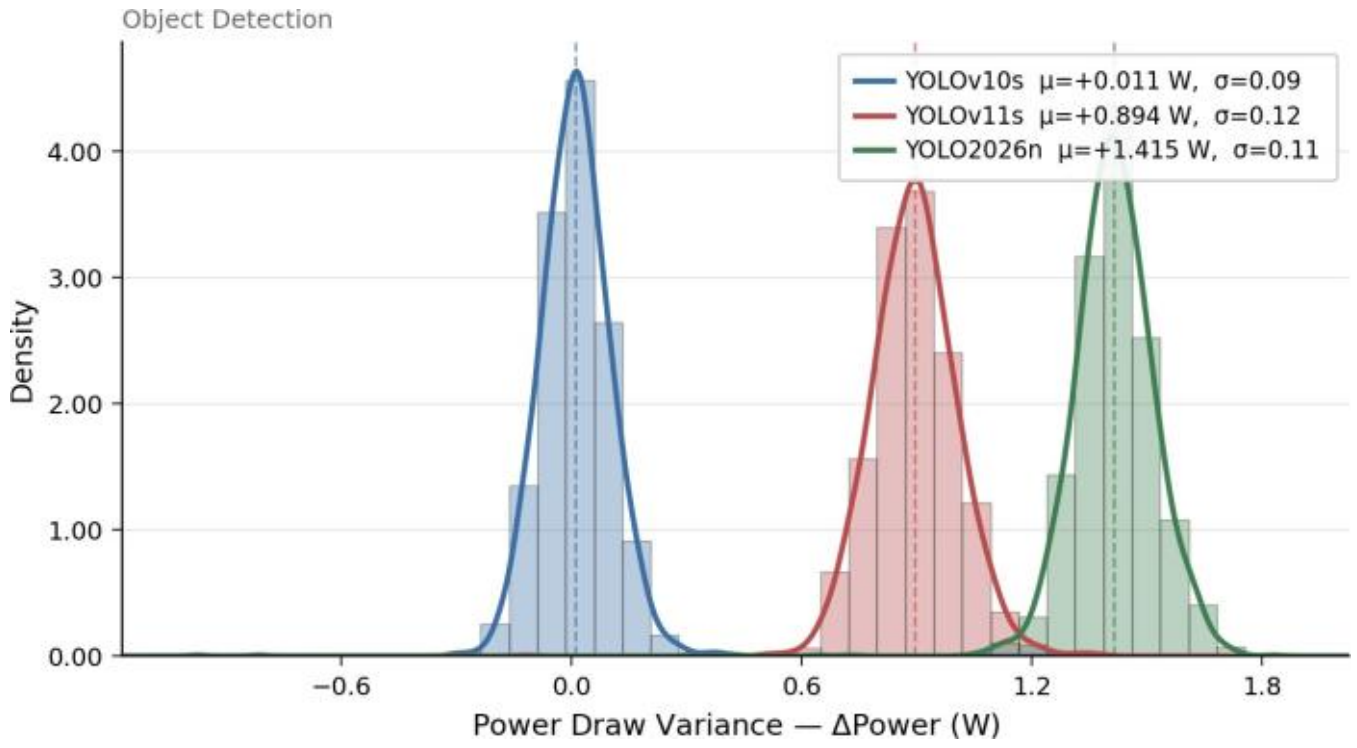

Fig. 5: Power Draw Variance ($\Delta$Power) for YOLO models under fault injection — object detection task.

*2) Memory Allocation (ΔRAM):* The `RAM_Delta` distributions show memory changes under fault injection for lane following and object detection (Figs. 6 and 7). All models exhibit negative shifts, indicating systematic memory release by the TensorRT engine relative to baseline.

YOLOv10s shows the largest release: lane following mean $-250.7$ MB ($\sigma = 71.4$ MB, band $[-387.4, +28.0]$ MB, absolute mean peak $2{,}648$ MB), object detection mean $-265.6$ MB ($\sigma = 128.0$ MB, band $[-450.0, +150.0]$ MB, peak $2{,}632$ MB), with a sharp unimodal distribution in lane following and broader variation for detection due to dynamic bounding-box allocation.

YOLOv11s exhibits the smallest release: lane following mean $-100.9$ MB ($\sigma = 40.9$ MB, band $[-227.7, +104.8]$ MB, peak $2{,}761$ MB), object detection mean $-117.1$ MB ($\sigma = 75.0$ MB, band $[-278.0, +166.5]$ MB, peak $2{,}745$ MB), with occasional secondary peaks near 0 MB for detection reflecting dynamic re-allocation.

YOLO2026n shows stable, task-invariant behaviour: lane following mean $-172.6$ MB ($\sigma = 75.8$ MB, band $[-355.7, +70.8]$ MB, peak $2{,}749$ MB), object detection mean $-168.2$ MB ($\sigma = 75.4$ MB, same band, peak $2{,}754$ MB), confirming consistent buffer management.

Overall, memory-release hierarchy is YOLOv10s ($\sim -258$ MB) $>$ YOLO2026n ($\sim -170$ MB) $>$ YOLOv11s ($\sim -109$ MB). All models remain within the Jetson Nano 4 GB ceiling under fault injection.

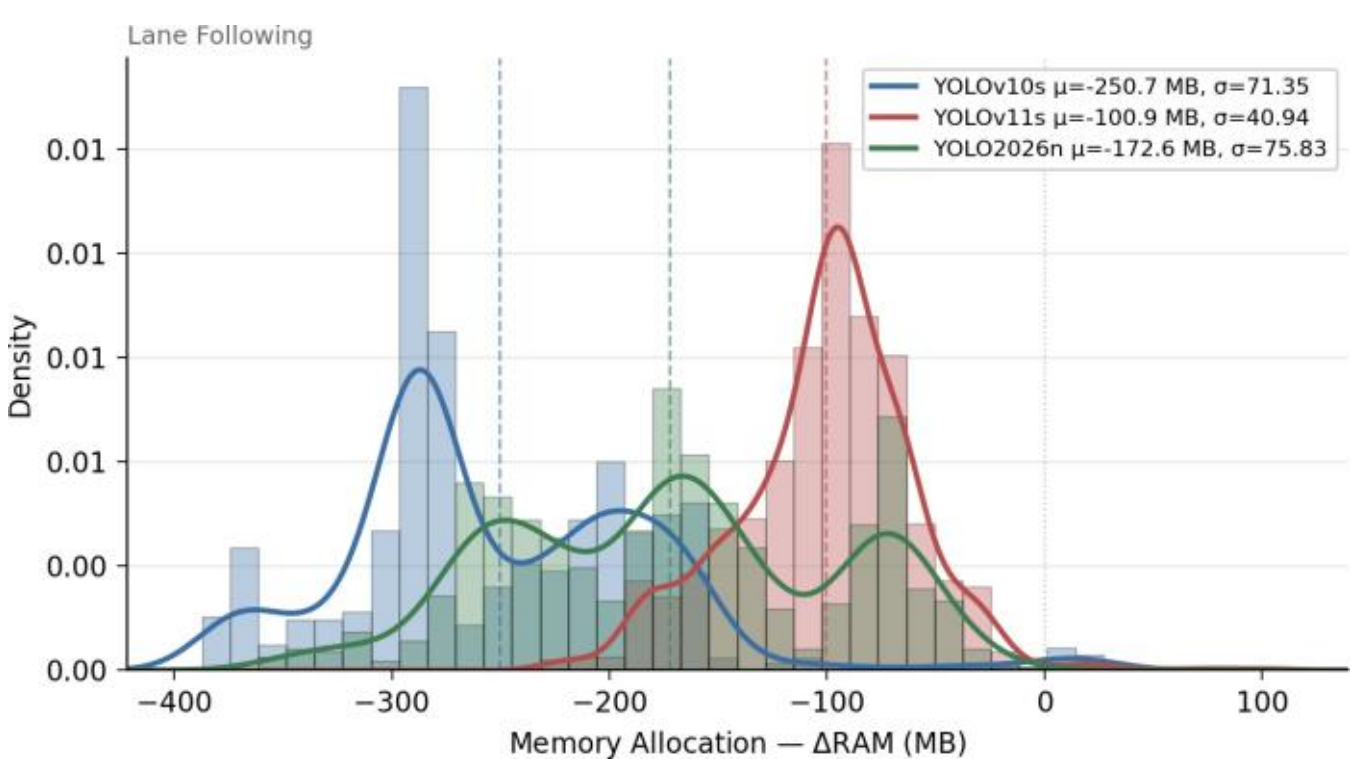

Fig. 6: Memory Allocation (ΔRAM) for YOLO models under fault injection — lane-following task.

*3) Thermal Flux (ΔTemp):* The `Temp_Delta` distributions show CPU temperature rise under fault injection (Figs. 8 and 9). All models produce strictly positive deltas in both tasks.

YOLOv10s exhibits the lowest increase: lane following mean $+2.414$ °C ($\sigma = 1.335$ °C, band $[-1.9, +6.5]$ °C, absolute mean $37.95$ °C), object detection mean $+4.550$ °C ($\sigma = 2.455$ °C, band $[-1.8, +10.0]$ °C, absolute mean $40.11$ °C).

YOLOv11s shows moderate rise in lane following and the highest in object detection: $+4.053$ °C ($\sigma = 1.385$ °C, band $[+0.3, +7.4]$ °C, absolute $40.36$ °C) and $+6.209$ °C ($\sigma = 2.593$ °C, band $[-0.4, +13.7]$ °C, absolute $42.52$ °C).

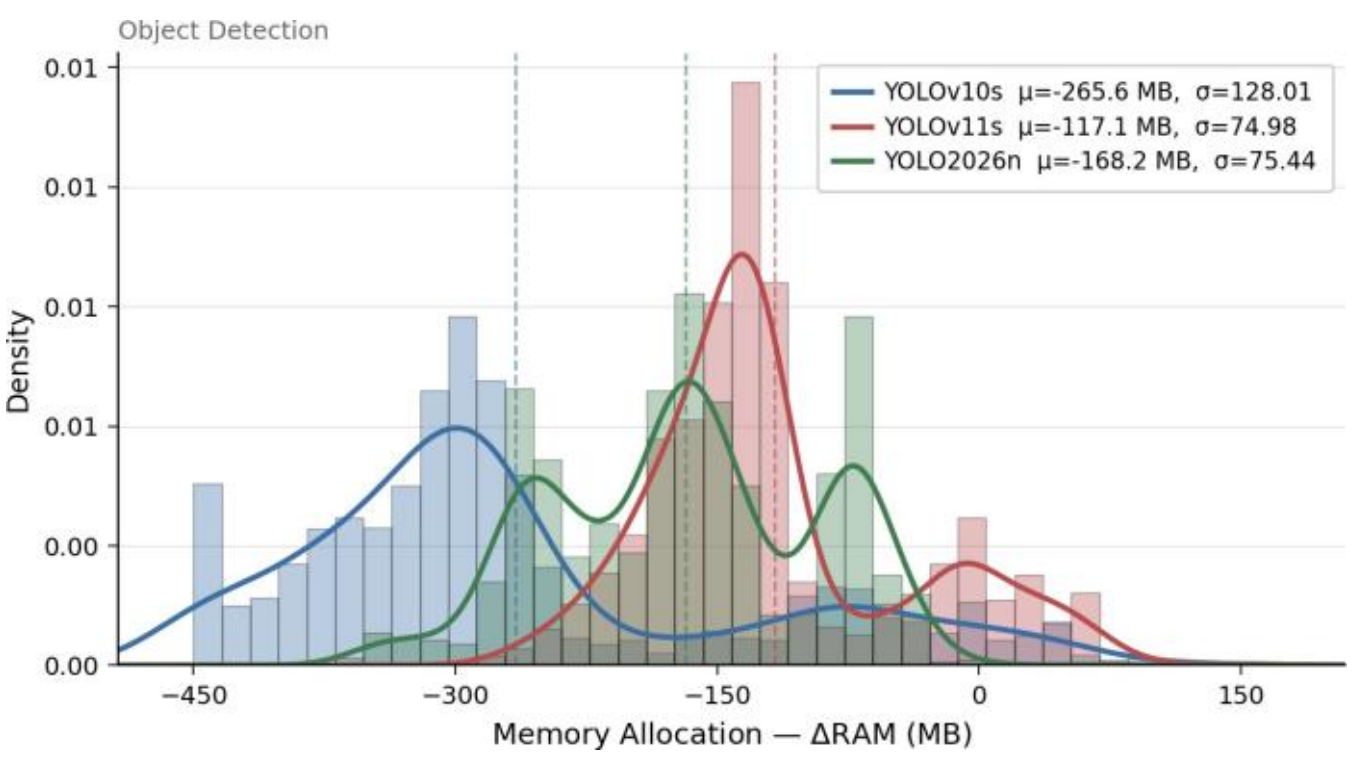

Fig. 7: Memory Allocation (ΔRAM) for YOLO models under fault injection — object detection task.

YOLO2026n is consistent across tasks: lane following $+5.943$ °C ($\sigma = 1.169$ °C, band $[+1.4, +8.6]$ °C, absolute $39.01$ °C), object detection $+5.939$ °C ($\sigma = 1.174$ °C, same band and absolute $39.01$ °C), confirming task-invariant thermal behaviour.

Thermal hierarchy across tasks: YOLOv10s ($+2.4$ °C LF, $+4.6$ °C OD) $<$ YOLO2026n ($\approx +5.9$ °C both) $<$ YOLOv11s ($+4.1$ °C LF, $+6.2$ °C OD). All models remain well below the Jetson Nano's throttling threshold.

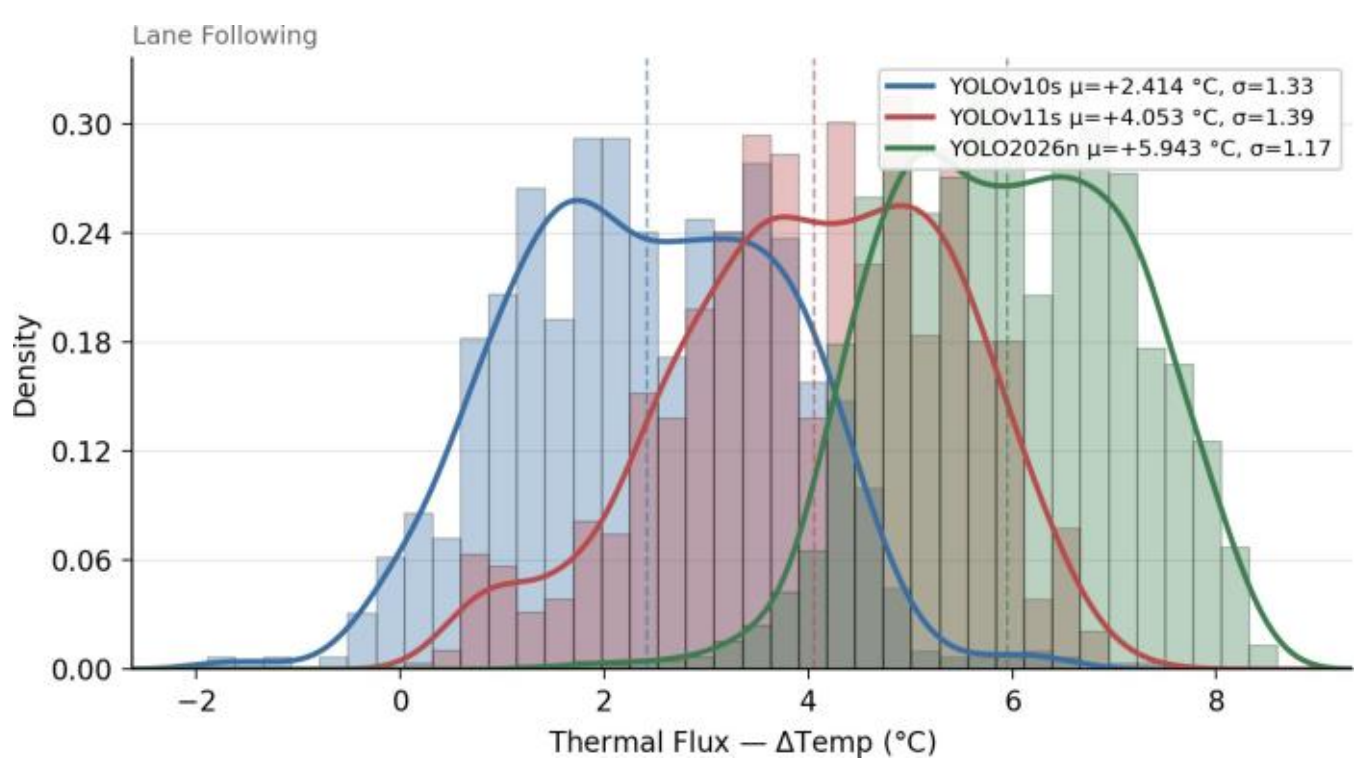

Fig. 8: Thermal Flux (ΔTemp) for YOLO models under fault injection — lane-following task.

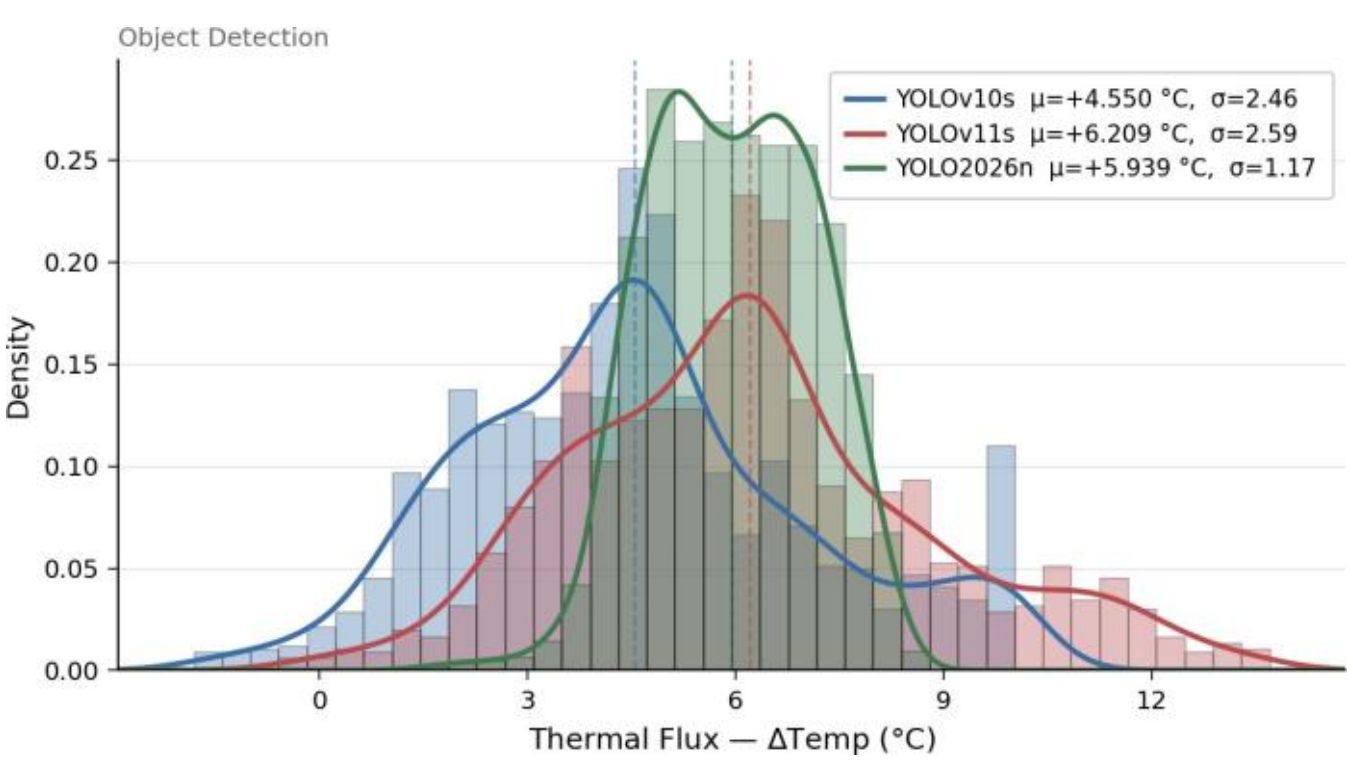

Fig. 9: Thermal Flux (ΔTemp) for YOLO models under fault injection — object detection task.

*4) Core Utilization (GPU%):* The `GPU_%` distributions show GPU occupancy across all fault-injection scenarios (Figs. 10 and 11). All models exhibit near-identical, bell-shaped distributions centered around 23–24%, confirming that GPU usage is hardware-bound rather than task- or model-bound.

YOLOv10s has the broadest distribution: lane following mean 23.8% ($\sigma = 6.95\%$, range [2.7, 40.9]%), object detection mean 23.7% ($\sigma = 6.91\%$, range [2.5, 48.0]%). YOLOv11s is the tightest: lane following 23.1% ($\sigma = 5.60\%$, range [7.2, 39.9]%), object detection 23.0% ($\sigma = 5.56\%$, range [5.8, 42.5]%). YOLO2026n is consistent across tasks: mean 23.2% ($\sigma \approx 5.8\%$, range [5.3, 43.4]%).

All distributions are symmetric with no skew or bimodality, indicating stable GPU scheduling. Maximum utilization remains below 44%, leaving ample headroom on the Jetson Nano, and all models operate comfortably under extended fault-injection load.

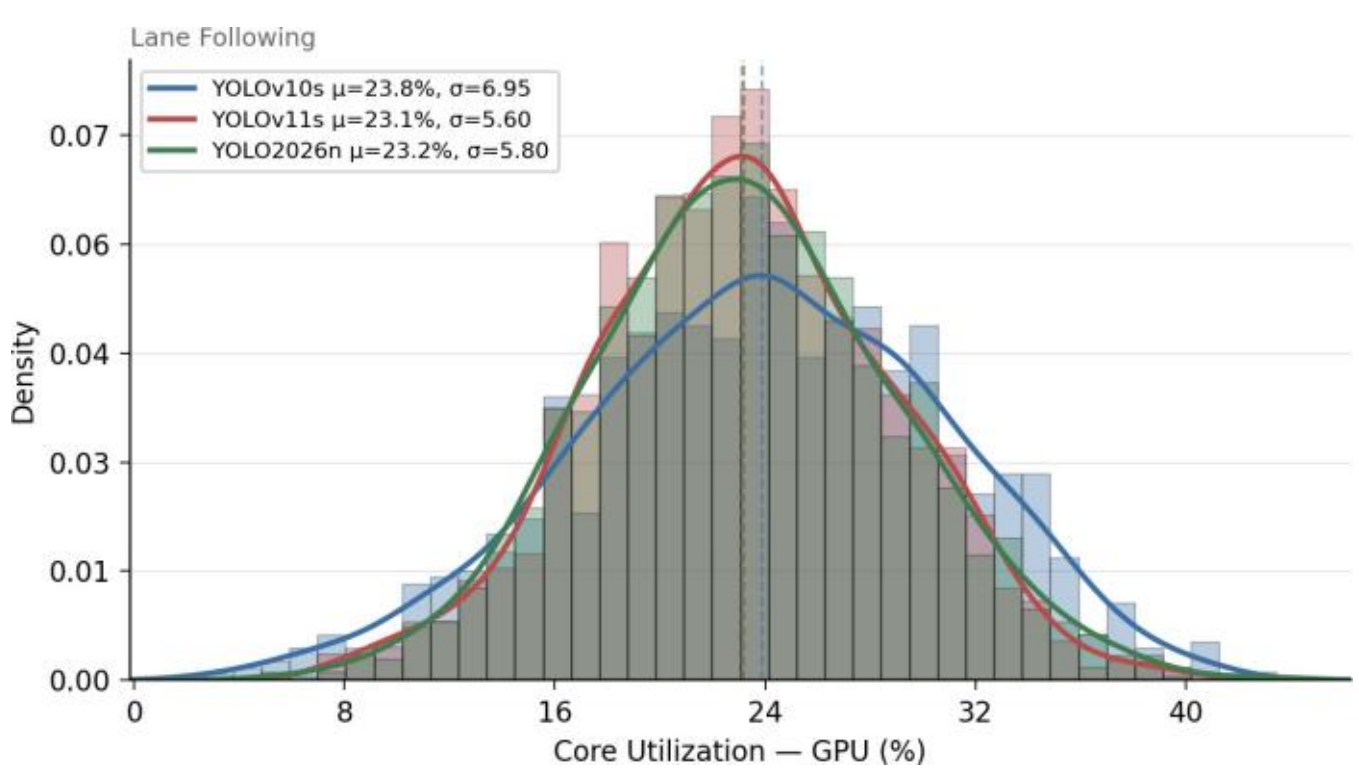

Fig. 10: Core Utilization (GPU%) for YOLO models under fault injection — lane-following task.

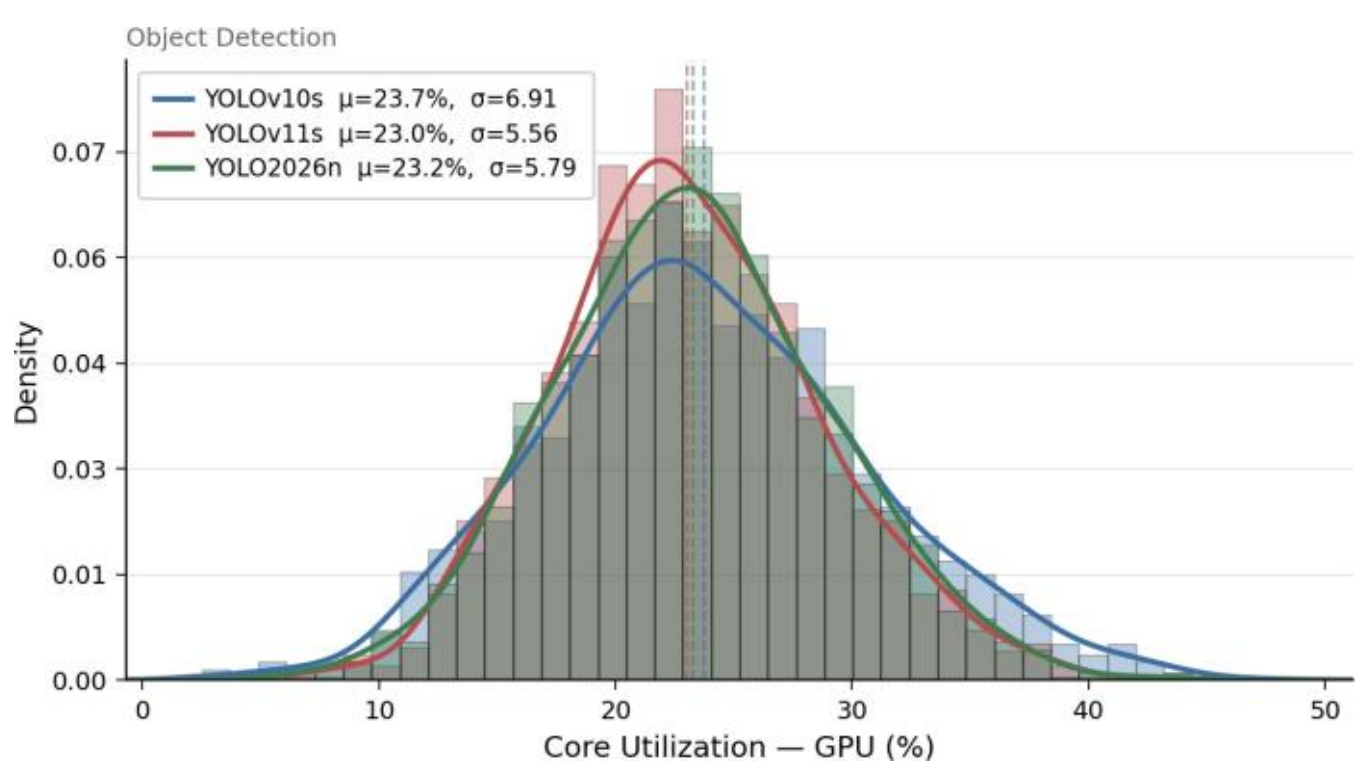

Fig. 11: Core Utilization (GPU%) for YOLO models under fault injection — object detection task.

## B. Inference Performance

The inference performance distributions, including Object Detection Stability (Retention%) and Throughput Variance ($\Delta$FPS), for object detection and lane following are shown respectively, measured across all fault-injected folders on the NVIDIA Jetson Nano.

*1) Detection Stability (Retention%):* The `Retention_%` distributions (Figs. 12 and 13) measure the fraction of baseline objects correctly detected under all fault folders. All models show left-skewed distributions concentrated in [88, 99]%, confirming robust detection under fault injection.

YOLOv10s has the lowest retention: lane following mean 89.4% ($\sigma = 4.64\%$, range [70.9, 97.1]%), object detection mean 89.1% ($\sigma = 4.91\%$, range [70.2, 98.6]%), with higher variability in object detection. YOLOv11s improves retention: lane following 91.8% ($\sigma = 3.86\%$, range [74.7, 98.8]%), object detection 91.5% ($\sigma = 4.09\%$, range [75.9, 99.4]%). YOLO2026n shows the highest and most consistent retention: 92.4% ($\sigma = 4.12\%$) in both tasks.

Overall hierarchy: YOLOv10s ($\approx$ 89.2%) $<$ YOLOv11s ($\approx$ 91.7%) $<$ YOLO2026n ($\approx$ 92.4%). All models retain $>$ 70% of detections even under worst-case faults.

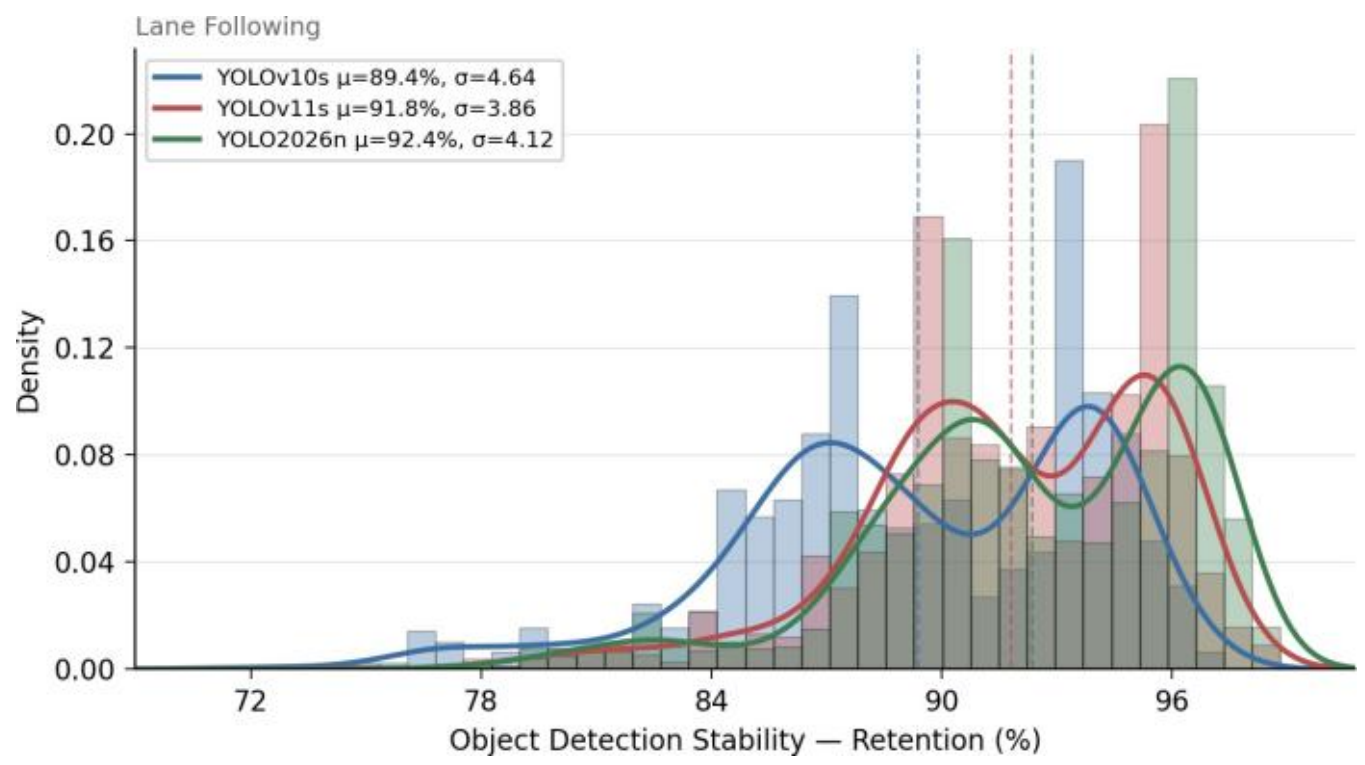

Fig. 12: Detection Stability (Retention%) for YOLO models under fault injection — lane-following task.

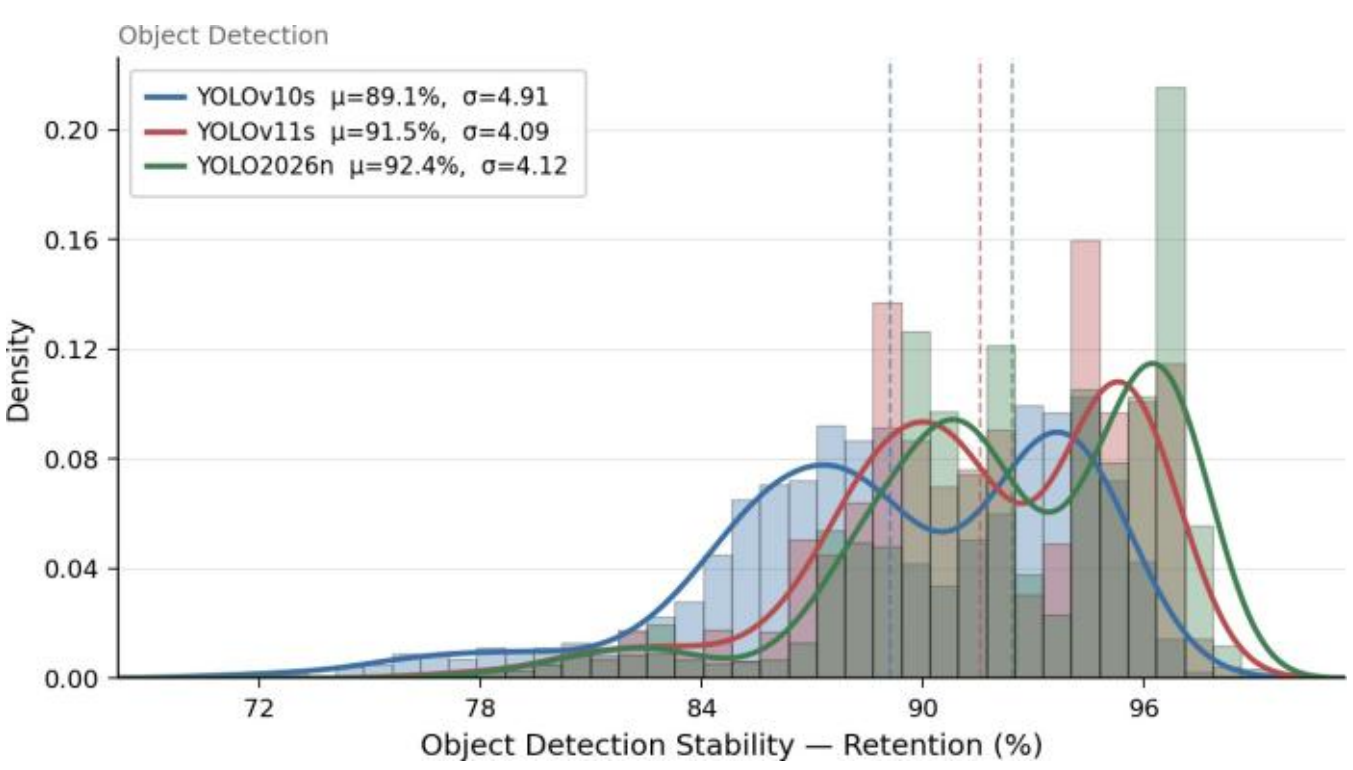

Fig. 13: Detection Stability (Retention%) for YOLO models under fault injection — object detection task.

*2) Throughput Variance ($\Delta$FPS):* The `FPS_Delta` distributions (Figs. 15 and 14) show the change in frames per second relative to baseline. Across all models and tasks, the mean $\Delta$FPS is positive, indicating that the TensorRT engine maintains or slightly improves throughput after the initial warm-up.

YOLOv10s exhibits a left tail down to $\sim -4$ FPS, while the majority of runs remain near or above baseline, confirming stable execution under fault injection.

s processed.

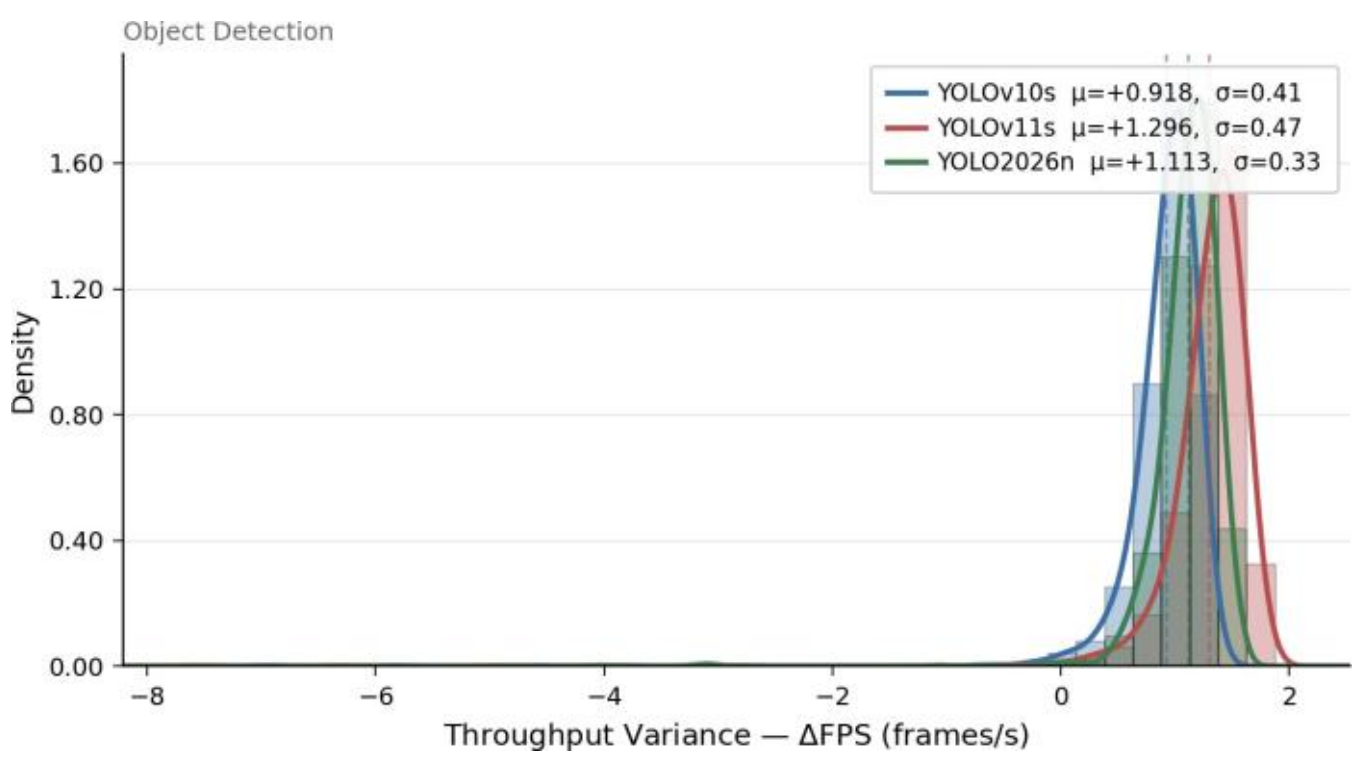

Fig. 14: Throughput Variance (ΔFPS) for YOLO models under fault injection — object detection task.

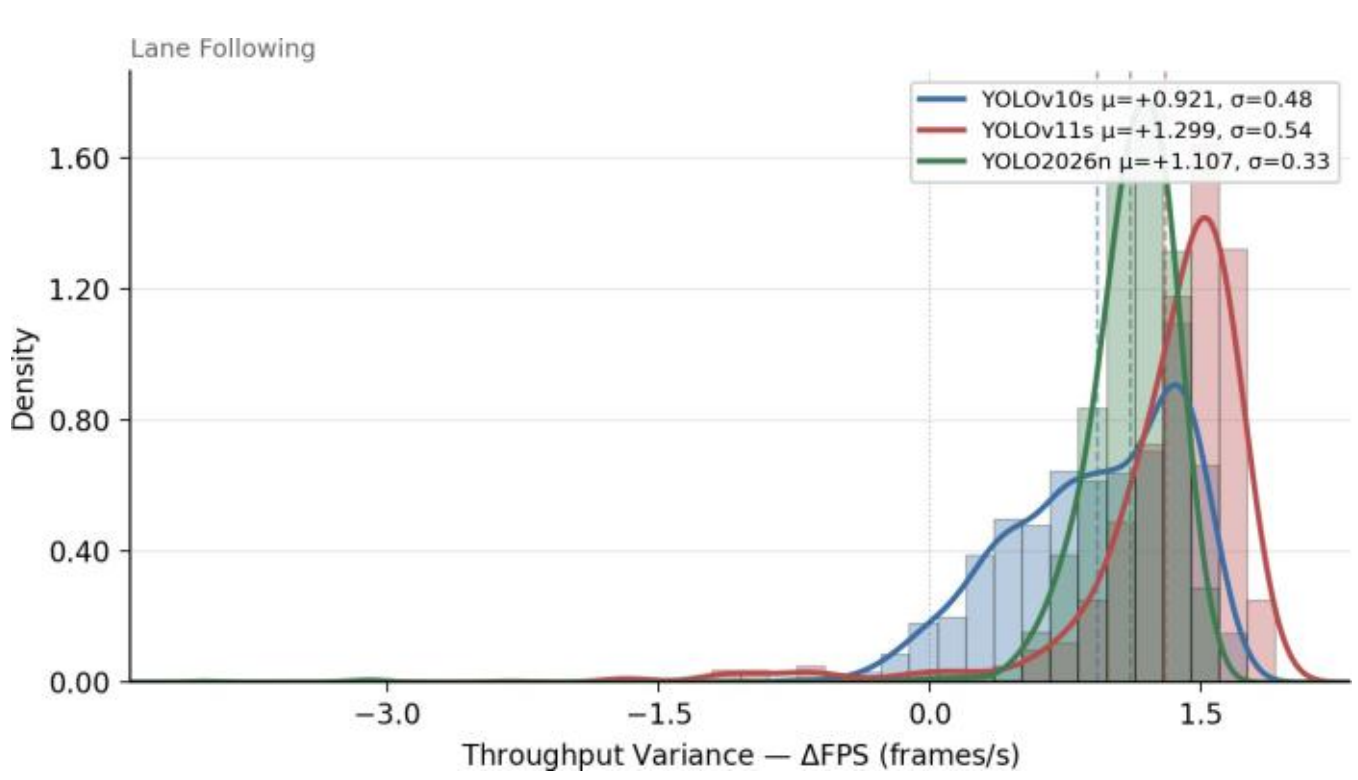

Fig. 15: Throughput Variance (ΔFPS) for YOLO models under fault injection — lane-following task.

## C. *Latency and Timing*

We illustrate the latency and timing distributions, including Tail Latency (Δp95), Tail Latency (Δp99), and Execution Consistency (ΔJitter), for object detection and lane following respectively, measured across all fault-injected folders on the NVIDIA Jetson Nano.

*1) Tail Latency (Δp95 and Δp99):* The `p95_Delta` and `p99_Delta` distributions (Figs. 16 and 17) show changes in 95th/99th percentile inference latency relative to baseline. YOLOv10s has positive p95 means in both tasks, while YOLOv11s and YOLO2026n have negative p95 means, indicating faster tail frames under fault injection.

YOLOv10s: lane-following mean p95 $+1.798$ ms ($\sigma = 2.339$, range $[-1.79, +5.22]$ ms), object detection $+1.627$ ms ($\sigma = 1.861$, range $[-0.13, +16.48]$ ms). p99 near zero: lane-following $+0.127$ ms, object detection $+0.092$ ms.

YOLOv11s: p95 lane-following $-0.700$ ms, object detection $-0.872$ ms; p99 lane-following $-1.096$ ms, object detection $-1.131$ ms. Occasional spikes up to +25 ms appear under high-severity faults.

YOLO2026n: p95 lane-following $-0.893$ ms, object detection $-0.913$ ms; p99 near zero ($+0.073$ and $+0.065$ ms), confirming tight task-invariant tail-latency.

Hierarchy (p95): YOLO2026n ($\approx 26.6$ ms) $\approx$ YOLOv11s ($\approx 27.0$ ms) $<$ YOLOv10s ($\approx 29.5$ ms). For p99: all $\approx 29.4$–$31.2$ ms, with isolated spikes.

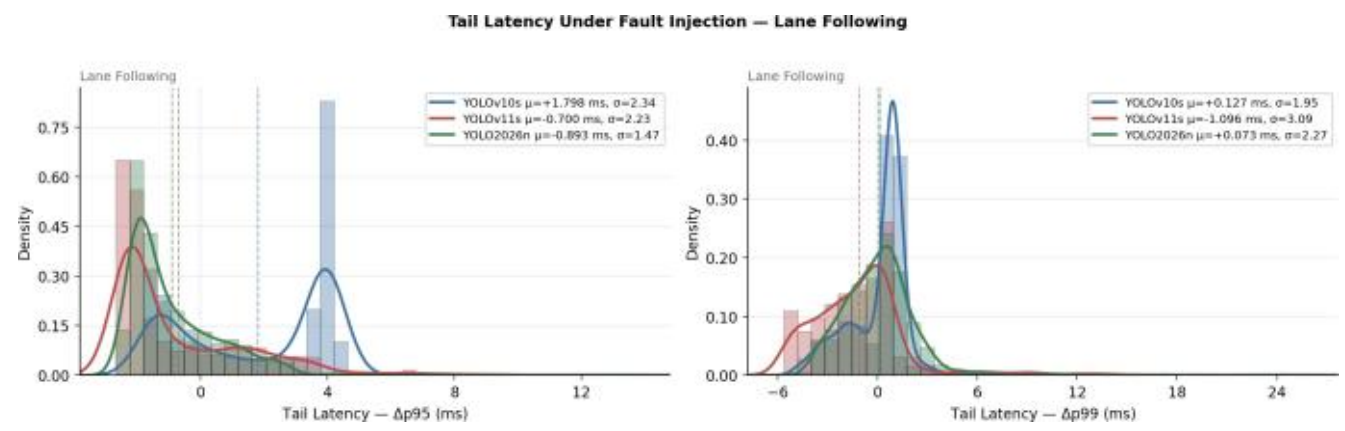

Fig. 16: Tail Latency (Δp95/p99) for YOLO models under fault injection — lane-following.

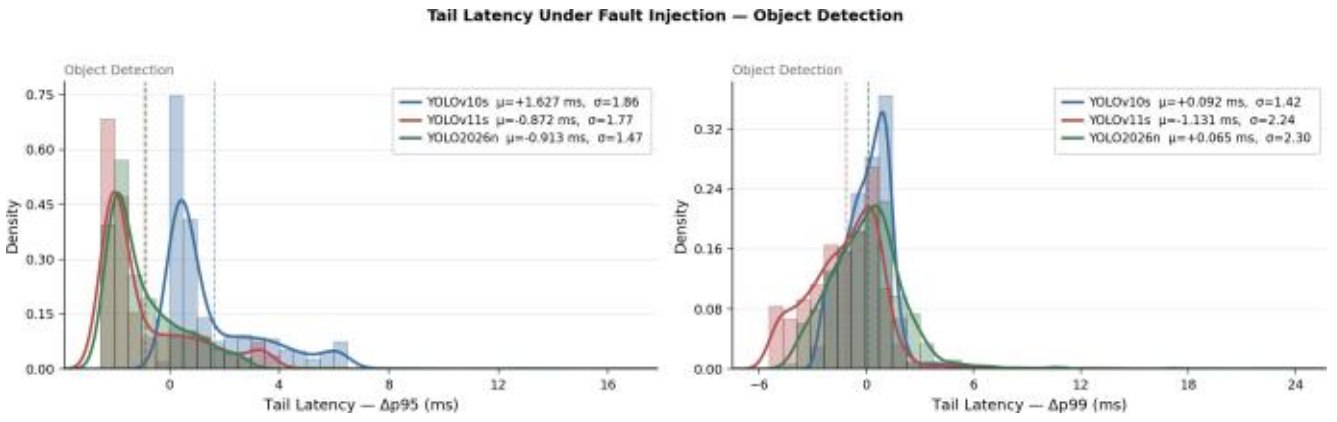

Fig. 17: Tail Latency (Δp95/p99) for YOLO models under fault injection — object detection.

*2) Execution Consistency (ΔJitter):* The `Jitter_Delta` distributions (Figs. 18 and 19) measure frame-to-frame timing variance. All distributions are right-skewed, centred near zero, with long tails up to $\sim 2.7$–$3.6$ ms. Most frames lie within $[0, 0.8]$ ms, indicating high consistency under fault injection.

YOLOv10s: lane-following mean $+0.102$ ms ($\sigma = 0.248$), object detection $+0.060$ ms ($\sigma = 0.146$). YOLOv11s: lane-following $+0.091$ ms ($\sigma = 0.426$), object detection $+0.049$ ms ($\sigma = 0.251$). YOLO2026n: lane-following $+0.087$ ms ($\sigma = 0.236$), object detection $+0.085$ ms ($\sigma = 0.239$), showing near-identical, task-invariant distributions.

All models maintain mean ΔJitter $< +0.11$ ms and absolute jitter well below 1 ms, confirming highly deterministic and rhythmically stable execution suitable for real-time deployment.

## D. *Cross-Model Comparison*

Table I compares hardware and inference metrics across the three YOLO variants in both lane following (LF) and object detection (OD).

YOLOv10s is the most power-efficient, adding almost no extra power, and releases the most memory. However, it shows higher latency and lower detection stability. YOLOv11s offers a balanced profile, with moderate power and thermal increase, while improving both throughput and latency compared to YOLOv10s. YOLO2026n delivers the best overall performance: highest detection stability, fastest and most consistent inference, and lowest tail latency, at the cost of higher power consumption. Across all models, GPU utilisation remains similar (∼23%), indicating hardware-bound execution with available headroom, and thermal levels stay below throttling limits. Overall, YOLOv10s is best for efficiency, YOLOv11s provides a good trade-off, and YOLO2026n is the strongest choice for stable, safety-critical edge deployment.

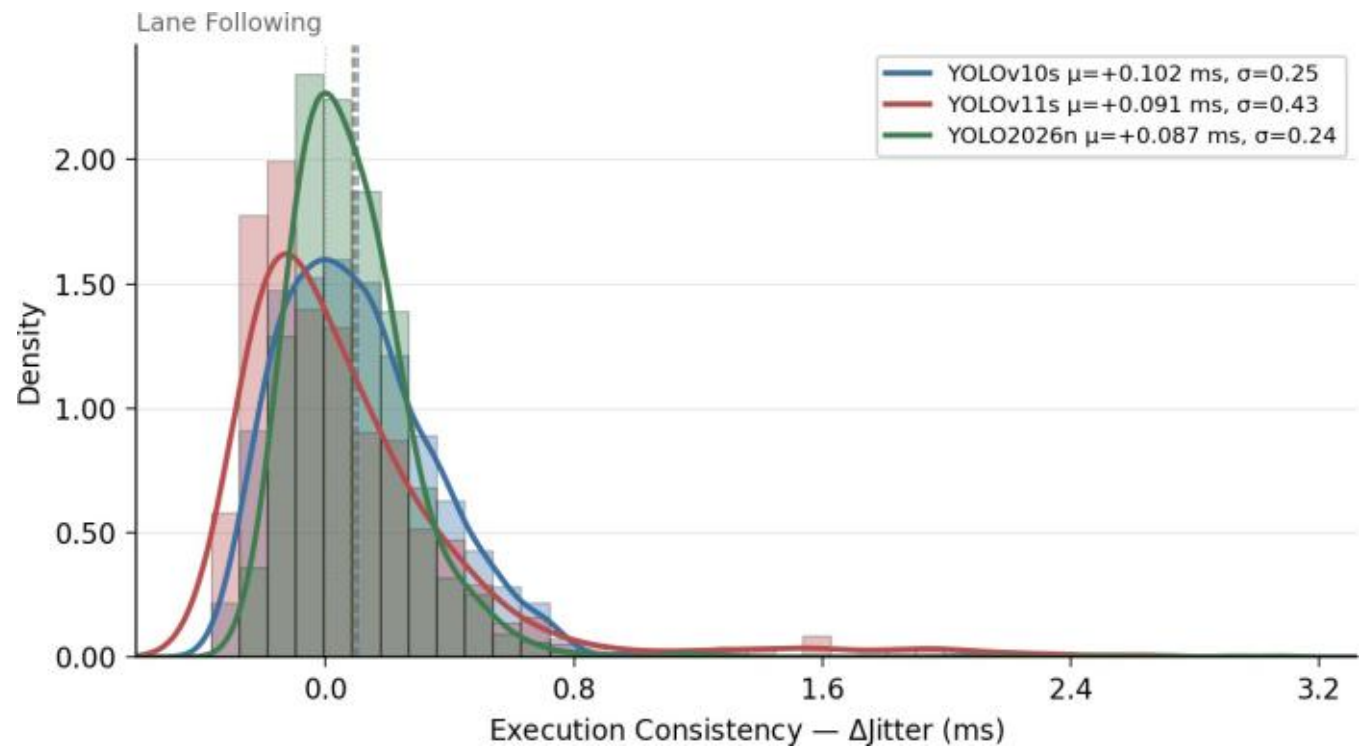

Fig. 18: Execution Consistency (ΔJitter) for YOLO models under fault injection — lane-following.

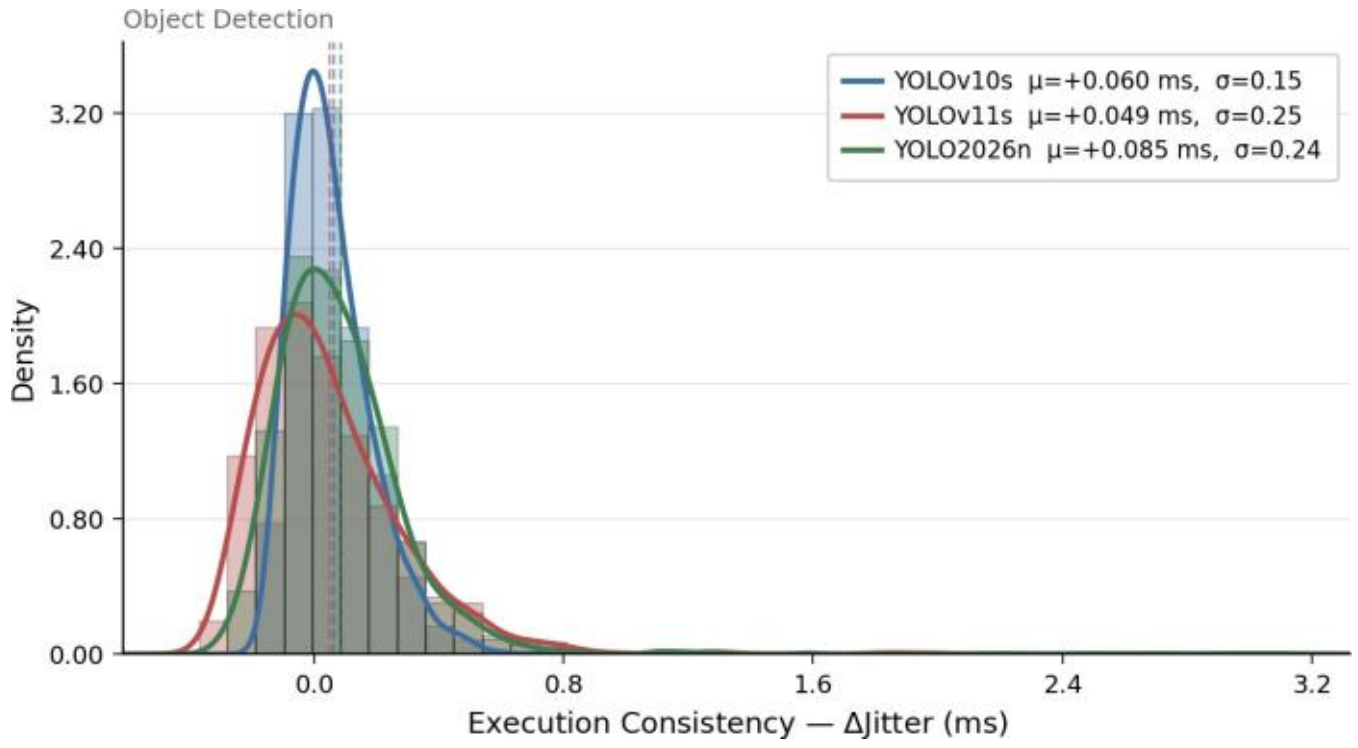

Fig. 19: Execution Consistency (ΔJitter) for YOLO models under fault injection — object detection.

## VII. Conclusion

This work evaluated YOLOv10s, YOLOv11s, and YOLO2026n deployed on an NVIDIA Jetson Nano under the VisionFault fault-injection taxonomy across lane-following and object detection tasks. All three TensorRT-optimised models maintain stable execution throughout: GPU utilisation converges to 23–24% regardless of model or task, memory usage decreases after warm-up (ΔRAM = $-109$ to $-266$ MB), thermal rise stays well below the throttling threshold (max $+6.2$ °C), and all models sustain real-time throughput with positive mean ΔFPS across both tasks.

The three models occupy distinct positions in the efficiency–stability trade-off. YOLOv10s is the most power-efficient (ΔPower $\approx$ 0 W) but has the highest tail latency (Δp95 $\approx$ $+1.7$ ms) and lowest retention ($\approx$89.2%). YOLOv11s achieves the highest throughput gain (ΔFPS $=$ $+1.30$) and most favourable tail latency (Δp95 $\approx$ $-0.88$ ms) at moderate power cost ($\approx$ $+$ 0.88 W). YOLO2026n delivers the best retention (92.4%), lowest and most stable tail latency (p95 $\approx$ 26.6 ms, $\sigma \approx 1.47$ ms), and the most task-invariant behaviour across all metrics, at the highest power draw ($\approx$ $+$ 1.41 W). For safety-critical edge deployment, YOLO2026n offers the best overall balance of robustness, latency predictability, and execution consistency. Future work will target fault-aware fine-tuning strategies for the identified worst-case fault categories.

TABLE I: Cross-Model Hardware and Inference Metrics — Mean Values Across Lane Following (LF) and Object Detection (OD)

| Metric | Task | YOLOv10s | YOLOv11s | YOLO2026n |
|---|---|---|---|---|
| ΔPower (W) | LF | −0.01 | +0.87 | +1.41 |
| | OD | +0.01 | +0.89 | +1.41 |
| ΔRAM (MB) | LF | −250.7 | −100.9 | −172.6 |
| | OD | −265.6 | −117.1 | −168.2 |
| ΔTemp (°C) | LF | +2.41 | +4.05 | +5.94 |
| | OD | +4.55 | +6.21 | +5.94 |
| GPU (%) | LF | 23.8 | 23.1 | 23.2 |
| | OD | 23.7 | 23.0 | 23.2 |
| Retention (%) | LF | 89.4 | 91.8 | 92.4 |
| | OD | 89.1 | 91.5 | 92.4 |
| ΔFPS | LF | +0.92 | +1.30 | +1.11 |
| | OD | +0.92 | +1.30 | +1.11 |
| Δp95 (ms) | LF | +1.80 | −0.70 | −0.89 |
| | OD | +1.63 | −0.87 | −0.91 |
| Δp99 (ms) | LF | +0.13 | −1.10 | +0.07 |
| | OD | +0.09 | −1.13 | +0.07 |
| ΔJitter (ms) | LF | +0.10 | +0.09 | +0.09 |
| | OD | +0.06 | +0.05 | +0.09 |